\documentclass{egpubl}

\JournalSubmission

\makeatletter
\copyrightTextTitPag{}
\copyrightTextRunPag{}
\def\ps@titlepage{\let\@mkboth\@gobbletwo
  \def\@oddhead{}%
  \def\@oddfoot{}%
  \let\@evenhead=\@oddhead
  \let\@evenfoot=\@oddfoot
  \let\sectionmark=\EmptySectionmark
  \let\subsectionmark=\EmptySubsectionmark}
\def\@oddfoot{}
\def\@evenfoot{}
\makeatother

\usepackage[T1]{fontenc}
\usepackage{dfadobe}
\usepackage{amsmath}
\usepackage{amssymb}
\usepackage{amsfonts}
\usepackage{float}
\usepackage{placeins}

\usepackage{cite}
\BibtexOrBiblatex

\electronicVersion
\PrintedOrElectronic
\ifpdf \usepackage[pdftex]{graphicx} \pdfcompresslevel=9
\else \usepackage[dvips]{graphicx} \fi

\usepackage{egweblnk}

\title[Volumetric Light Transport Inference]%
      {Real-Time Volumetric Light Transport Inference \\ without a Path}

\author[L. Leonard \& R. Westermann]
{\parbox{\textwidth}{\centering L.\,Leonard$^{1}$
        and R.\,Westermann$^{1}$
        }
        \\
{\parbox{\textwidth}{\centering $^1$Technical University of Munich, Germany\\
       }
}
}

\begin{document}
\teaser{
\includegraphics[width=1.0\linewidth]{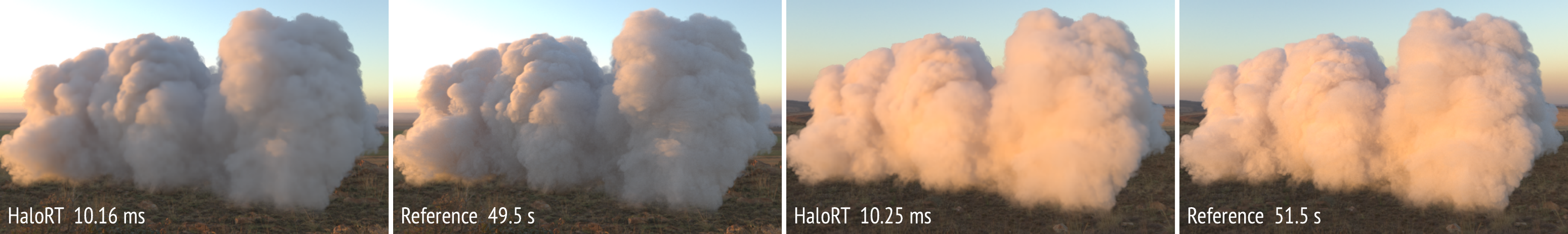}
 \centering
  \caption{For clouds and environments it has never seen during training, our attention-based neural model reconstructs full volumetric global illumination from inexpensive primary-ray features and environment samples in real-time on an NVIDIA GeForce RTX 5090. The computation of path-traced radiance samples is entirely avoided. 
  }
\label{fig:teaser}
}

\maketitle
%-------------------------------------------------------------------------
\begin{abstract} 
We present a real-time method for reconstructing fully integrated pixel radiance in rendered volumetric media under direct and ambient environmental illumination, using compact, low-cost scene features and statistics. Unlike Monte Carlo denoising pipelines for volumetric rendering, which depend on computationally expensive path-traced radiance estimates, our approach directly infers the final light-transport solution from inexpensive primary-ray and environment sample evaluations. We show that these features form a near-sufficient representation of the scene, enabling accurate reconstruction at substantially lower computational cost. Using an attention-augmented image-to-image network, we demonstrate that complete pixel radiance can be inferred directly from these auxiliary features, without requiring the original radiance signal. Operating in image space also provides a more stable computational cost as the camera zooms into or out of the volume, with substantially less dependence on volume density than approaches based on explicit multiple-scattering simulation. We demonstrate real-time performance and high reconstruction accuracy in challenging volumetric cloud scenarios.
%-------------------------------------------------------------------------
\end{abstract}  
%-------------------------------------------------------------------------
\section{Introduction}

\begin{figure*}[t]
    \centering
    \includegraphics[width=\linewidth]{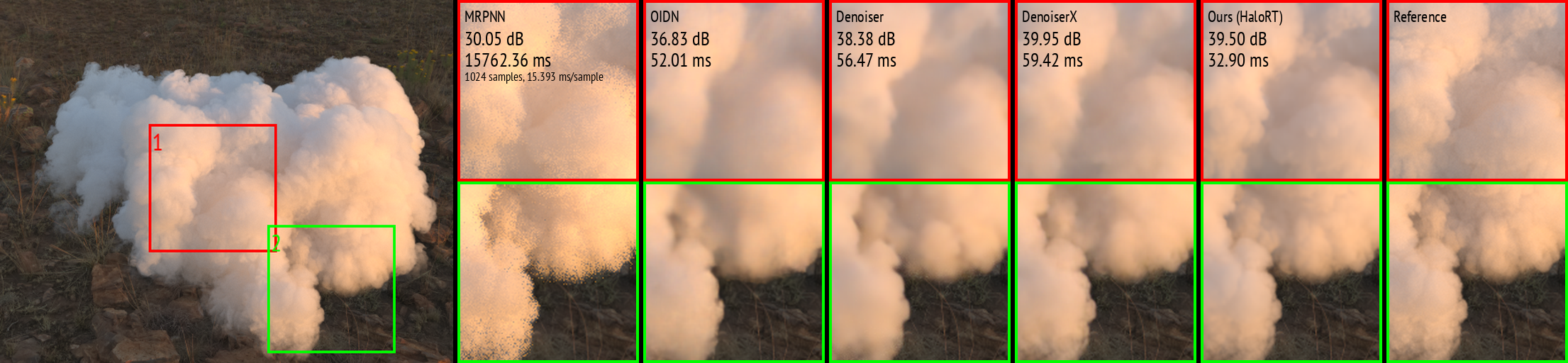}
    \caption{Comparison under realistic ambient environment lighting. A density-conditioned transport estimation \cite{Hu2023MRPNN} (MRPNN) uniformly samples the environment and produces noise even with 1024 samples; Intel Open Image Denoise \cite{OpenImageDenoise} (OIDN) uses only rendered pixel color and smooths out fine details; a specialized volumetric denoiser \cite{hofmann2021volumerecon} (Denoiser) better preserves appearance but suffers from stochastic noise; a variant of the volumetric denoiser using our deterministic features (DenoiserX) better preserves fine details. All denoisers use $4$ radiance samples per pixel. Ours, without any path samples, achieves superior detail reconstruction at lower computational cost. All methods run on an NVIDIA GeForce RTX 3090, as MRPNN is incompatible with newer GPU architectures.}
    \label{fig:sota_comparison}
\end{figure*}

Monte Carlo (MC) path tracing achieves photorealistic rendering by simulating the full light transport pipeline, including emission, scattering, and absorption along importance-sampled light paths~\cite{kajiya1986rendering,cook1986stochastic,arvo1990particle,novak2018monte,pharr2023physically}.
For surface rendering, light paths terminate after a few bounces, yet each pixel still requires hundreds to thousands of samples to converge to a noise-free result. For volumetric media, the problem is substantially more severe: light scatters at every point along a ray, turning each sample into a recursive integration problem. Rendering a single frame of participating media such as clouds, smoke, or fog often requires seconds to minutes even on high-end GPUs, making real-time applications prohibitive.

To mitigate the cost of path tracing, many modern rendering systems rely on MC denoising. For surface rendering, denoisers leverage auxiliary features---often called G-buffers in surface graphics---like albedo, normals, depth, and variance buffers to filter sparse path samples (given as samples per pixel, or spp) into smooth outputs~\cite{bako2017kernel,vogels2018denoising,schied2017spatiotemporal}, or to use them as input for neural networks that predict noise-free images directly \cite{chaitanya2017interactive,isik2021interactive,Vavilala-2024-DiffDenoise}.

These MC denoisers must evaluate at least a few full light transport paths to accurately transform the noisy rendering into a high-quality image. This becomes especially costly for volumetric media, where simulating light transport is computationally expensive. Moreover, paths in volumetric media exhibit complex, view-dependent correlations across samples that surface-based auxiliary features cannot exploit, rendering such features less predictive (Fig.~\ref{fig:sota_comparison} OIDN). Due to the nature of participating media, the noise distribution itself varies spatially and spectrally, further complicating inference.

One approach to addressing these challenges is MC denoising guided by auxiliary features specifically designed to represent volumetric light transport characteristics, such as transmittance and first-scatter event estimates \cite{Hofmann2020neural, hofmann2021volumerecon}.
While effective in principle, such methods still require path-traced radiance samples as input: their cost grows with medium density, and the noise of these samples propagates into the reconstruction, leading to blurring and instability across realizations (Fig.~\ref{fig:sota_comparison} Denoiser).

In another research line, radiance-predicting neural networks replace path tracing for atmospheric clouds, albeit at considerably higher computational cost than MC denoisers \cite{kallweit2017deep} or requiring progressive inference updates to simulate full environment lighting \cite{Hu2023MRPNN} (Fig.~\ref{fig:sota_comparison}).

Both lines of work still depend on simulating light transport through the volume at inference time, either through path samples or through per-sample network queries. This raises a simple question: \emph{are path samples needed at all?} Prior work on surfaces shows that localized illumination effects can be inferred from auxiliary buffers alone~\cite{nalbach2017deep}, and Deep Scattering shows that networks can learn volumetric transport~\cite{kallweit2017deep}. We build upon both insights: we keep the feature-based design of volumetric denoisers, but drop the radiance input entirely and regress the final radiance from features computed along primary rays.

We show that optical depth, multi-scale interaction depths, transmittance towards the light, environment illumination at the expected first interaction, and local medium properties form a near-sufficient representation of the transport solution. We formalize this as two testable assumptions, sufficiency and noise invariance, and learn the mapping from features to radiance with an attention-augmented image-to-image network. Since the features are deterministic except for low-variance light-visibility estimates, the output is stable across realizations without requiring accumulation over frames.

We demonstrate real-time performance and high reconstruction accuracy in challenging volumetric cloud scenes (Fig.~\ref{fig:teaser} and Fig.~\ref{fig:sota_comparison} Ours), where state-of-the-art MC denoisers struggle to maintain both stability and fine detail. By moving from sample denoising to primary-ray feature regression, our method offers a new pathway toward practical real-time volumetric light transport.

In summary, our contributions are:
\begin{itemize}
    \item The formulation of volumetric rendering as a deterministic decoding from inexpensive transport features, replacing path sampling at inference time.
    \item A compact set of primary-ray features for participating media, which we show to be near-sufficient for inferring full multiple scattering in clouds.
    \item A real-time image-to-image network with local self-attention, which infers directional and environmental scattering separately and scales to high resolutions.
    \item A real-time cloud renderer simulating full radiance transport in unseen clouds and environments, showing high accuracy, low variance across realizations, and stable cost even for high densities.
\end{itemize}

\section{Related Work}

\emph{MC denoising for surfaces.}
Early denoisers handle the spatially varying noise of MC renderings with nonlinear image filters, such as bilateral and edge-avoiding filters~\cite{lee1983digital,rushmeier1994energy,dammertz2010edge} or non-local means~\cite{Rousselle2012Adaptive}, often guided by G-buffers to preserve edges. Spatiotemporal variance-guided filtering~\cite{schied2017spatiotemporal} additionally reuses samples from past frames to improve temporal continuity. With the advent of deep learning, denoising has shifted from handcrafted filters to data-driven prediction~\cite{Huo2021MonteCarloDenoisingSurvey,yan2025survey}. Learning the mapping from noisy features to filter parameters~\cite{kalantari2015machine} gave rise to two paradigms: radiance-prediction networks regress clean pixel values directly~\cite{chaitanya2017interactive,xu2019adversarial,lu2020dmcr}, whereas kernel-prediction networks estimate spatially varying filter kernels that keep the output consistent with the input colors~\cite{bako2017kernel,vogels2018denoising,isik2021interactive}. Sampling and temporal denoising networks can further be trained jointly~\cite{hasselgren2020neural}. These methods achieve near-reference quality by treating rendering as conditional image synthesis, but they are designed for surfaces and all require path-traced radiance as input.

\emph{Volumetric denoising and sample reuse.}
Volumetric path tracing converges far more slowly than its surface counterpart, since each radiance sample involves many scattering events along the ray. At low sample counts this yields severely noisy images, while auxiliary buffers that are effective for surfaces, such as depth or normals, become noisy or ill-defined in media~\cite{yan2025survey,Denisova-2025a}. Hofmann et al.~\cite{Hofmann2020neural} were the first to extend neural MC denoising to volumetric path tracing of medical data, introducing a set of auxiliary features designed specifically for volumes and increasing the effective sample count through spatial and temporal filtering; this was later extended to sparse volumes~\cite{hofmann2021volumerecon}. Our features build on this idea. Subsequent work decomposes the image into a surface and a volume layer with dedicated denoisers and a network that learns to blend both~\cite{Hofmann2023Joint}. Xu et al.~\cite{Xu2025} decouple radiance samples according to their noise level and fuse them with auxiliary features in a lightweight network, reducing the bias of earlier decoupling strategies. In contrast, Iglesias-Guitian et al.~\cite{iglesias2022realtime} explicitly avoid auxiliary features and exploit temporal coherence instead, reprojecting linear models across frames and combining past and current estimates by weighted recursive least squares. Orthogonally, the quality of the input samples can be improved. Zhou et al.~\cite{ZHOU201797} fit Gaussian mixtures to single-scattered radiance to guide visibility-aware sampling in participating media, and spatiotemporal reservoir resampling~\cite{Bitterli-2020Restir} has been extended to volumetric path space by ReSTIR-Volume~\cite{lin2021fast}. All these approaches still trace paths through the medium; our method requires none at inference time.

\emph{Radiance-free neural rendering.}
Concurrently with learned denoisers, Deep Shading~\cite{nalbach2017deep} showed that CNNs can synthesize the final appearance directly from screen-space G-buffers, circumventing hand-crafted shaders and requiring no radiance input at all. Subsequent work explicitly frames G-buffers as the input representation for learning light transport, positioning G-buffer-based neural rendering as a modern alternative to precomputed radiance transfer~\cite{NPRT-Rainer2022}. G-buffers have since been combined with reflective shadow maps to reconstruct complex indirect lighting~\cite{kim2026lightweight}, used to synthesize global illumination including transparent objects~\cite{zhang2026neural}, and shown to remain valuable even when the primary transport data comes from other sources~\cite{Huang2022RealtimeDR}. Beyond screen space, transformer-based models infer global illumination directly from a 3D scene description: RenderFormer~\cite{zeng2025renderformer} maps triangle tokens with their reflectance properties to images with full global illumination, and Xu et al.~\cite{xu2026generalizable} learn a generalizable light transport embedding from scene point clouds, queried through cross-attention to predict diffuse global illumination in unseen scenes. These approaches recover local and global illumination effects on surfaces, but none addresses the direct inference of global illumination in volumetric media. We transfer their central idea, radiance-free inference from auxiliary buffers, to participating media.

\emph{Learned volumetric light transport.}
Instead of denoising rendered images, neural networks can approximate components of volumetric light transport itself. Deep Scattering~\cite{kallweit2017deep} predicts the radiance caused by multiple scattering in clouds from a hierarchical 3D descriptor of the density around each shading point, oriented towards the light source. Learned statistical path models replace long stochastic scattering chains with a surrogate based on a more structured probabilistic model of the local geometry~\cite{Ludwig-2021}. MRPNN~\cite{Hu2023MRPNN} brings Deep Scattering to real time with a lightweight multi-feature fusion network that combines precomputed transmittance fields with a frequency-sensitive density stencil. For environment illumination, however, it requires progressive sampling over light directions, repeating the full ray accumulation for each direction. Rather than evaluating a network per sample, Liang et al.~\cite{liang2026multifeature} bake a compact low-rank latent approximation of the scattering operator, gaining performance and avoiding over-illumination artifacts, but at the cost of a baking step per volume. Similarly, Relightable Neural Assets~\cite{mullia2024rna} encode all shading and scattering of an asset in a feature grid decoded by a small MLP, an approach adapted to relight volumetric clouds under environment illumination within a rasterization pipeline~\cite{olajos2026environmental}; such representations are tied to the specific volume they were trained on. In contrast, our network operates in image space on features computed on the fly, requires neither per-point 3D descriptors nor precomputed volumetric fields or per-volume training, and generalizes to unseen volumes.

\begin{figure*}[t]
    \includegraphics[width=\linewidth]{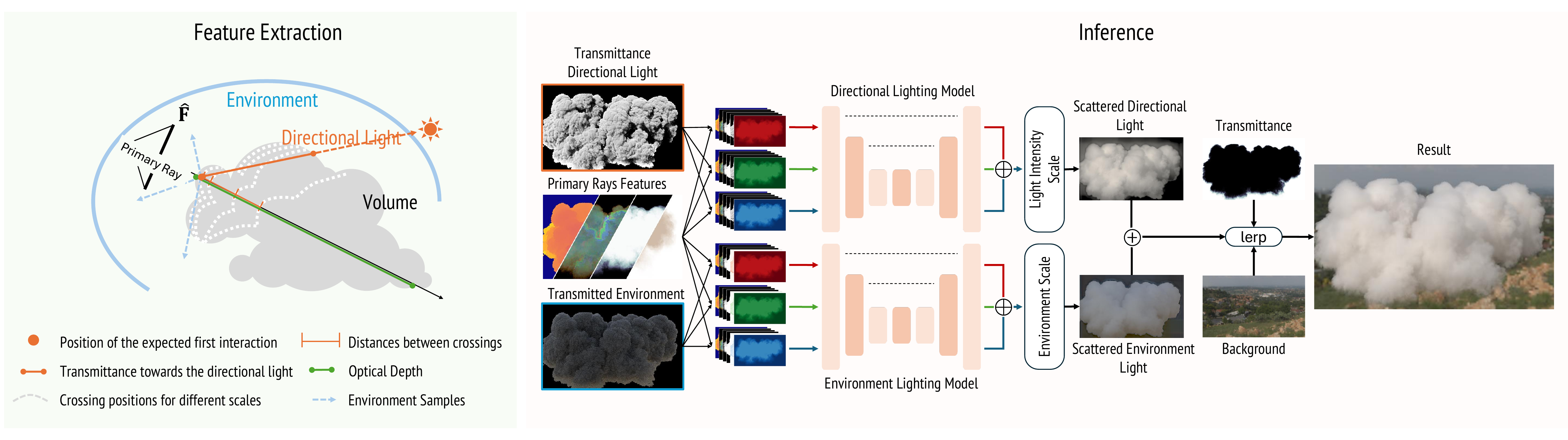}
    \caption{Feature extraction and light-transport inference.
Primary rays are converted into compact features describing optical depth, multi-scale density crossings, transmittance toward directional and environmental illumination, and local scattering properties.
These features are organized into light-transport channels and processed by specialized neural models to infer the corresponding scattered-light components, which are subsequently combined into the final image.}
    \label{fig:full_process}
\end{figure*}

\section{Method}

Fig.~\ref{fig:full_process} gives an overview of our method. We first formulate rendering as a deterministic decoding of image features (Secs.~\ref{sec:pbvr}--\ref{sec:formulation}), then define a set of inexpensive primary-ray and environment features for optically thick media (Sec.~\ref{sec:features_vr}), and finally describe how the decoder is learned and structured to separate directional and environmental light (Secs.~\ref{sec:learning_decoder}--\ref{sec:structured_decoder}).

\subsection{Physically-Based Volume Rendering}
\label{sec:pbvr}

Let $\theta$ denote a scene parameterization including geometry, materials, illumination, and participating media. We define a physically-based rendering operator $R$ such that $I = R(\theta)$ corresponds to the solution of the light transport equation. Evaluating $R$ is computationally expensive due to high-dimensional path integrals required for global illumination with multiple scattering.

Light transport in participating media is governed by the volume rendering equation, which accounts for attenuation, absorption, emission, and multiple scattering along a ray. Let $x_t = x_0 - t\omega$ denote a point along direction $-\omega$ from an origin $x_0$. The outgoing radiance is given by
\[
L(x_0,\omega)
=
\int_0^d
T(x_0 \leftrightarrow x_t)\,
\sigma_t(x_t)\,
\bigl[\cdots\bigr]\,
dt
+
T(x_0 \leftrightarrow x_d)\,
L(x_d,\omega),
\]
where $\sigma_t$ is the extinction coefficient and
\[
T(x_a \leftrightarrow x_b)
=
\exp\!\left(
-\int_{x_a}^{x_b} \sigma_t(x)\,dx
\right)
\]
denotes transmittance between two points in the medium.

The term $[\cdots]$ represents the local source contribution,
\[
\varphi(x_t)\,L_s(x_t,\omega)
+
\bigl(1-\varphi(x_t)\bigr)\,L_e(x_t,\omega),
\]
where $\varphi$ is the single-scattering albedo, $L_e$ denotes emitted radiance, and $L_s$ is the in-scattered radiance. The latter integrates incident illumination over the sphere of directions:
\[
L_s(x_t,\omega)
=
\int_{\mathcal{S}^2}
\rho(x_t,\omega,\omega_i)\,
L(x_t,\omega_i)\,
d\omega_i,
\]
where $\rho$ is the phase function describing angular scattering. In this paper, media with emissive particles are not considered, i.e., $L_e(x,\omega)=0$.
We employ the Henyey--Greenstein phase function, parameterized by an anisotropy coefficient $g$, which provides an efficient approximation of highly forward-scattering volumetric transport.

\subsection{Problem Formulation}
\label{sec:formulation}

We introduce a set of auxiliary forward operators $\{F_k\}_{k=1}^n$ producing intermediate images
\[
\hat{\mathbf{F}} = \mathbf{F}(\theta;\xi) = \bigl(F_1(\theta;\xi),\dots,F_n(\theta;\xi)\bigr),
\qquad \xi \sim p(\xi),
\]
where $\xi$ denotes the random numbers consumed by the operators, drawn independently of the scene $\theta$. Unlike $R$, each $F_k$ is a computationally efficient transport projection derived from $\theta$, encoding structured quantities such as transmittance, scattering integrals, or low-order illumination terms rather than approximations of full radiance.

We make the following assumptions:

\textbf{(A1) Sufficiency.}
The rendered image depends on the scene only through the distribution of its features:
\[
\mathbf{F}(\theta;\xi) \overset{d}{=} \mathbf{F}(\theta';\xi) \;\Rightarrow\; R(\theta) = R(\theta'),
\]
where $\overset{d}{=}$ denotes equality in distribution with respect to $\xi \sim p(\xi)$. Hence, the scene parameters $\theta$ are not required to infer the image. Notably, (A1) does not require $\theta$ itself to be recoverable from the features: different scenes, e.g., different density fields, may share the same feature distribution, as long as they also share the same rendered image.

\textbf{(A2) Noise-invariant decoding.}
There exists a deterministic mapping $S$ that is insensitive to the stochastic estimation of the features, i.e., all realizations of the features of a scene decode to the same image:
\[
S\bigl(\mathbf{F}(\theta;\xi)\bigr) = S\bigl(\mathbf{F}(\theta;\xi')\bigr) = R(\theta)
\qquad \text{almost surely}.
\]

Under (A1) and (A2), rendering reduces to a deterministic mapping from noisy auxiliary features to radiance. Both assumptions can be tested separately: violations of (A1) appear as a systematic bias that persists even when the decoder observes arbitrarily many realizations $\mathbf{F}(\theta;\xi_1), \mathbf{F}(\theta;\xi_2), \dots$ of the features, i.e., when their distribution is fully known, while violations of (A2) appear as a non-zero variance $\mathrm{Var}_{\xi}\bigl[S(\mathbf{F}(\theta;\xi))\bigr]$ across single realizations.

This formulation generalizes MC denoising, where a single operator $F_{\mathrm{MC}}(\theta;\xi)$ provides an unbiased radiance estimate. Its distribution determines $R(\theta) = \mathbb{E}_{\xi}[F_{\mathrm{MC}}(\theta;\xi)]$, so (A1) holds trivially, and the entire difficulty lies in (A2).
Our operators shift this balance: most features are deterministic and the remaining ones have low variance, making (A2) easy to satisfy, while none of them is a degraded version of $I$. 
Instead, each $F_k$ provides a complementary, physically meaningful projection of light transport, and $S$ performs radiance reconstruction from heterogeneous transport measurements rather than denoising of a single stochastic estimator.

\subsection{Features}
\label{sec:features_vr}

We now define the operators $\{F_k\}$ for volume rendering in optically thick media. All of them are evaluated along primary rays and describe attenuation, visibility, illumination, and local scattering properties (Fig.~\ref{fig:full_process} Left).

\emph{Optical depth $F_\tau$.}
We consider the accumulated extinction for each ray
\[
\tau(d)=\int_0^d \sigma_t(x_t)\,dt,
\]
which governs transmittance, expected free-flight distance, and scattering density. Rather than using transmittance directly,
we employ a normalized encoding of a medium with reduced extinction
\[
\hat{\tau}(d)=1-\frac{1}{1+0.5\,\tau(d)}.
\]
This transformation suppresses the exponential saturation of $T$ in high-opacity regions, yielding a smoother signal that preserves relative variations in dense media and improves numerical conditioning during reconstruction.

\emph{Interaction depth $F_d$.}
We estimate the depth of the first scattering event, conditioned on an interaction occurring along the ray. Let $\tau=\tau(d)$, then
\[
\tau_c = 1 - \frac{\tau e^{-\tau}}{1-e^{-\tau}}
\]
is the expected optical depth of the first interaction, given that it occurs before $d$. We define the interaction depth as the minimum distance $t_1$ at which this optical depth is reached,
\[
\tau(t_1) = \tau_c.
\]
Unlike the expected distance, which may fall into empty space between separated structures, $t_1$ always lies inside the medium. It removes the noise and discontinuities of the free-flight estimator and is always defined for $\tau>0$.

\emph{Crossing layers $F_c$.}
To capture volumetric structure across scales, we compute multiple interaction-depth fields using progressively reduced extinction levels. Each layer represents the first conditioned interaction in a progressively more transparent version of the medium, producing a continuous multi-scale profile of volumetric occupancy.
Instead of using the depth sequence, we use the distance between two consecutive crossings. This value remains more stable to camera variations.

\emph{Transmittance for the next-event estimation $F_n$.}
For the first interaction point $x_{t_1}$, we estimate the transmittance towards a directional light source $T(x_{t_1} \leftrightarrow x_l)$.

\emph{Environment illumination $F_e$.}
We additionally compute incident illumination from the environment map convolved through transmittance and phase function:
\[
\int_{\mathcal{S}^2}
T(x_{t_1} \leftrightarrow x_i)\,
\rho(x_{t_1},\omega,\omega_i) L(x_i, \omega_i)\,
d\omega_i.
\]
We estimate this integral using a low-sample-count Monte Carlo estimator, with $8$ samples as a practical trade-off between computational cost and stability; this number can be adjusted as needed (see Section~\ref{sec:implementation}).

\emph{Medium properties $F_m$.}
At the first interaction point, we extract local scattering parameters including the Henyey--Greenstein anisotropy parameter $g$ and the single-scattering albedo $\varphi(x_{t_1})$.

Together, these auxiliary quantities define a structured representation of the medium that captures partial yet informative constraints on global light transport.

\subsection{Learning the Decoder}
\label{sec:learning_decoder}

Assumptions (A1) and (A2) postulate the existence of the decoder $S$, but provide no closed form for it. We therefore approximate $S$ by a parameterized model
\[
S_\Psi : \mathbb{R}^{H \times W \times n} \rightarrow \mathbb{R}^{H \times W \times 3},
\]
an image-to-image neural network with weights $\Psi$ that maps the $n$ feature images of resolution $H \times W$ to the rendered image. The parameters are obtained by minimizing the expected reconstruction error over both the distribution of scenes and the stochastic source of the features:
\[
\Psi^\star
=
\arg\min_{\Psi}\;
\mathbb{E}_{\theta \sim p(\theta)}\,
\mathbb{E}_{\xi \sim p(\xi)}
\Bigl[
\ell\bigl(S_\Psi(\mathbf{F}(\theta;\xi)),\, R(\theta)\bigr)
\Bigr],
\]
where $\ell$ is a reconstruction loss. In practice, the outer expectation is approximated by a finite set of training scenes $\theta_i$ with converged reference renderings $R(\theta_i)$, and the inner expectation by drawing several realizations $\xi$ for each scene during training.

Both expectations relate directly to our assumptions. The inner expectation pairs all realizations $\mathbf{F}(\theta;\xi)$ of a scene with the same target $R(\theta)$, and thus rewards outputs that are invariant to $\xi$, as required by (A2). The outer expectation restricts the sufficiency assumption (A1) to the scene distribution $p(\theta)$: $S_\Psi$ only needs to be accurate for plausible scenes, and may exploit regularities of $p(\theta)$, e.g., typical cloud structures, to resolve ambiguities left open by the features.

\subsection{Structured Decoder}
\label{sec:structured_decoder}

The formulation above holds for any decoder. In practice, we exploit exact properties of the rendering operator $R$ to reduce the complexity of the learning problem. Since these properties are satisfied by $R$ itself, the true decoder $S$ shares them, and imposing them on $S_\Psi$ restricts the hypothesis space without excluding the solution.

\emph{Linearity in the illumination.}
We split the distant illumination into a smooth environment map $L_{\mathrm{env}}$ and a dominant directional light $\omega_l$ with intensities $e$ and $l$. Since light transport is linear in the incident radiance,
\[
I = e\,\bigl(T\,B + I_e\bigr) + l\,I_l,
\]
where $T\,B$ is the environment transmitted along the primary ray, and $I_e$, $I_l$ are the radiance scattered towards the camera for a normalized environment and a unit directional light. $T$ and $B$ are known from the primary-ray traversal, so only $I_e$ and $I_l$ need to be inferred. We learn them with two decoders, $S_e$ and $S_l$, each receiving the shared primary-ray features and the illumination feature of its component ($F_e$ resp.\ $F_n$). Factoring out $e$ and $l$ also removes the illumination scale from the inputs, and a single prediction covers arbitrary intensity ratios (Fig.~\ref{fig:full_process} Right).

\emph{Independence across color channels.}
Without spectral coupling, each color channel is the solution of an independent transport problem. We therefore apply the same single-channel decoder to each channel, with the features of that channel as input. This shares weights across channels and reduces the output to a single channel. The decoder is agnostic to which medium properties vary across channels, since each channel is decoded from its own features. Which properties vary only affects the cost of feature extraction: a channel-dependent extinction requires a separate traversal of the volume per channel. In our experiments, we use a shared extinction and vary only the albedo and illumination per channel, so the density-dependent features are computed once and shared.

\emph{Decoupling by construction.}
A single model that jointly predicts all components and channels would have to learn these independences from data. This requires training cases that cover the combinations of illumination and channel-dependent medium properties, and additional losses that penalize the influence of one channel or component on the others. In our experiments, such joint models did not achieve the required independence. In addition, they required considerably more capacity, which reduced performance substantially. Separate single-channel decoders satisfy both properties by construction and keep each decoder small.

The specific architecture of the decoders, the loss $\ell$, and the training procedure are described in Sec.~\ref{sec:implementation}.

\section{Implementation}
\label{sec:implementation}

The rendering pipeline is implemented in Vulkan, with CUDA-exported device memory used to directly use rendered buffers in the neural network inference stage, avoiding host-side synchronization overhead and ensuring a tightly coupled rendering--inference loop.

\subsection{Feature Rendering}

Our scenes consist of volumes scaled to fit in a $[-1,1]^3$ cube. We employ exhaustive ray-marching (step size $0.001\mathrm{u}$) to remove stochasticity for the primary ray features. We do a DDA on the coarse grid of majorants to skip empty regions when traversing the volume, and ray-march only non-zero voxels. A first traversal computes the optical depth $\tau$. A second traversal starts from the first position with non-zero density, starting to determine the crossings, with scales $\times4,\times1,\times0.25, \times0.1$. Because density scales are sorted, we can handle all crossings with just a second traversal. Early termination is possible when all crossings have been determined.

For the first crossing depth, values are estimated for the transmittance towards the directional light, the environment samples, the anisotropy factor, and the single-scattering albedo.

To reduce variance in the transmitted direct-lighting estimate, we use deterministic ray marching over a short initial segment of radius $0.1\mathrm{u}$, followed by stochastic ratio tracking. For environment lighting, we importance-sample directions using a quadtree over the equirectangular environment map, with sampling densities proportional to luminance.

\subsection{Network Architecture}

\begin{figure}
    \centering
    \includegraphics[width=\linewidth]{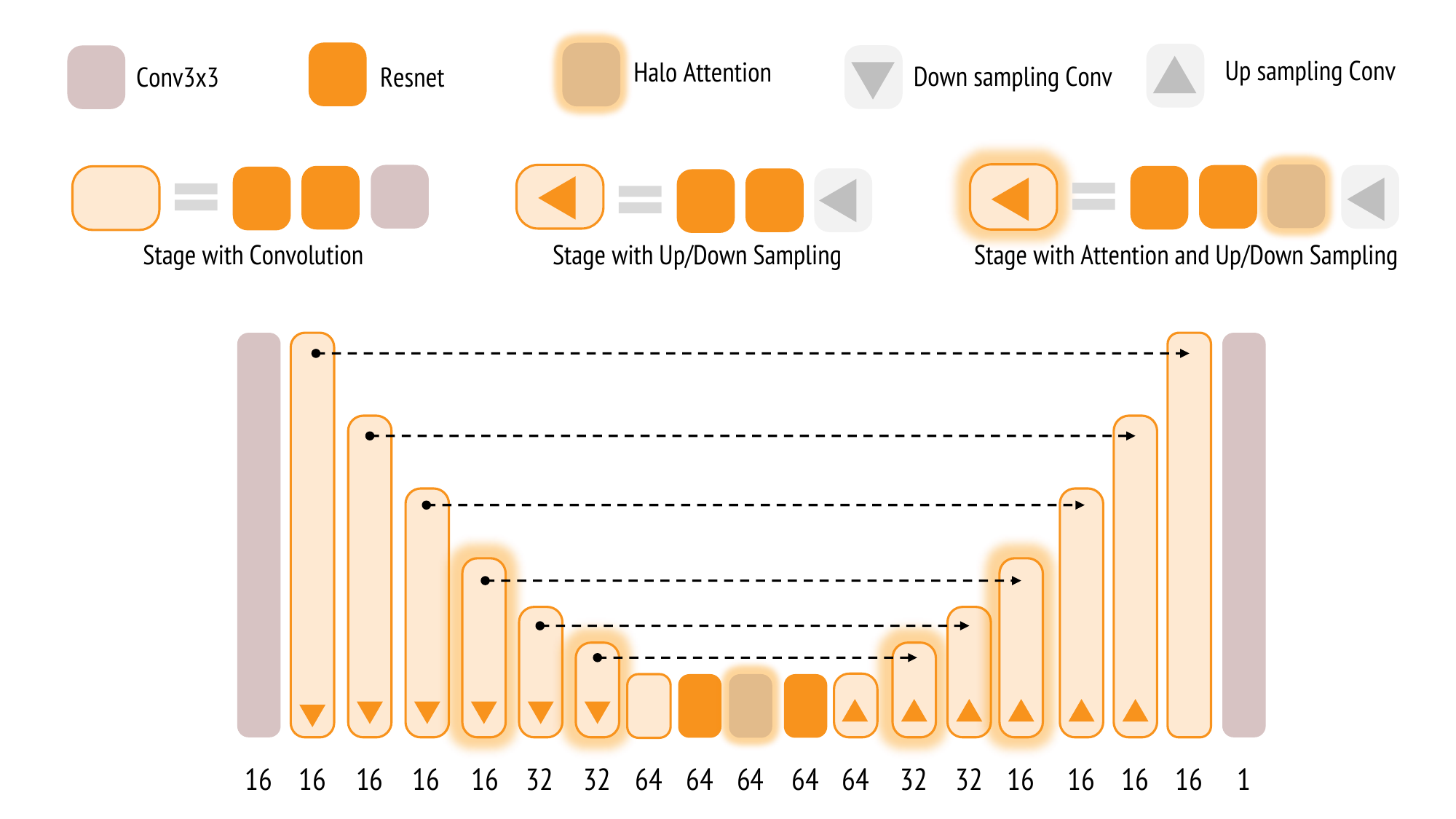}
    \caption{Architecture of our HaloRT model. Our U-Net downsamples $6$ times the initial resolution and works with blocks of $64$ channels in the bottleneck. Local window attention (Halo) is applied instead of full attention in 4 stages and the bottleneck.}
    \label{fig:halort}
\end{figure}

The reconstruction operator $S$ is parametrized using a U-Net backbone composed of residual blocks and attention-based feature mixing. Following HaloNet~\cite{vaswani2021scaling}, we incorporate local spatial self-attention at intermediate resolutions and at the bottleneck. We call our model \textbf{HaloRT}---\textbf{Halo}-attention-based \textbf{R}adiative \textbf{T}ransfer model (see Fig.~\ref{fig:halort}). 

Since multiple scattering couples distant regions of the medium, a pixel's radiance can depend on features at other image locations whose relevance varies with the scene, e.g., thin regions through which light enters the volume. Thus, we complement the convolutional backbone with local attention that relates features through content-dependent weights rather than fixed kernels (see Table~\ref{tab:arch_scaling}).

Restricting attention to local spatial blocks achieves similar scalability with image resolution as CNN-based models used in classical denoisers (4th row in Table~\ref{tab:arch_scaling}) and avoids the quadratic computational and memory costs of full spatial self-attention (3rd row of Table~\ref{tab:arch_scaling}).
At the bottleneck, the reduced spatial resolution still allows each local attention block to operate on features whose receptive fields span the entire input image, enabling image-wide information exchange at a small additional cost compared to attention-free inference (2nd row in Table~\ref{tab:arch_scaling}). 

\begin{table}
\centering
\caption{Scaling of inference-time (ms) with resolution for different 8-channel-in / 1-channel-out architectures, measured on an NVIDIA GeForce RTX 5090.}
\label{tab:arch_scaling}
\begin{tabular}{lrrrr}
\hline
Architecture & $256^2$ & $512^2$ & $1024^2$ & $2048^2$ \\
\hline
HaloRT & 1.411 & 1.563 & 2.914 & 10.528 \\
HaloRT without attention & 1.233 & 1.378 & 2.646 & 10.007 \\
Full attn at same stages & 1.446 & 1.618 & 4.700 & 31.659 \\
CNN-based Autoencoder & 1.029 & 1.132 & 1.835 & 6.830 \\
\hline
\end{tabular}
\end{table}

\begin{table*}[ht]
\centering
\begin{tabular}{lcccccc}
\hline
Method & MAE$\downarrow$ & PSNR$\uparrow$ (dB) & SSIM$\uparrow$ & LPIPS$\downarrow$ & RMS-SD$\downarrow$ & Time$\downarrow$ (ms) \\
\hline
Denoiser & 0.00391 $\pm$ 0.00179 & 41.62 $\pm$ 3.05 & 0.9687 $\pm$ 0.0250 & 0.0901 $\pm$ 0.0652 & 0.00628 $\pm$ 0.00330 & 23.84 $\pm$ 9.61 \\
DenoiserX & 0.00338 $\pm$ 0.00151 & \textbf{44.23 $\pm$ 2.80} & 0.9827 $\pm$ 0.0151 & 0.0517 $\pm$ 0.0420 & 0.00323 $\pm$ 0.00123 & 24.76 $\pm$ 9.98 \\
HaloRT & \textbf{0.00324 $\pm$ 0.00265} & 42.79 $\pm$ 3.91 & \textbf{0.9874 $\pm$ 0.0112} & \textbf{0.0420 $\pm$ 0.0418}
& \textbf{0.00070 $\pm$ 0.00035} & \textbf{10.66 $\pm$ 0.52} \\
\hline
\end{tabular}
\caption{Quality and performance comparison, averaged over the evaluation dataset. Our attention-based HaloRT achieves the best overall performance except for the PSNR, whereas our proposed DenoiserX remains competitive.}
\label{tab:metrics}
\end{table*}

Notably, the network is never explicitly provided with the light direction. Nevertheless, it can reconstruct backlit regions effectively by relating the directional light transmittance to the other extracted features, exploiting the spatially coherent visibility pattern of the light source across the image.

\subsection{Rendering and Dataset Generation}

We render volumetric scenes using delta-tracking combined with next-event estimation (NEE). Light transport is simulated through stochastic free-flight sampling and a single scattering evaluation along the ray. At each scattering event, direct illumination is estimated via explicit light sampling of the environment, while indirect contribution is accumulated through MC integration with a single sample over the phase function that continues estimating higher-order scattered light. Only for the primary rays, the case with no interaction accumulates the environment light (transmitted background).

We further use ratio-tracking to estimate the transmittance, in combination with a coarse grid of majorants to accelerate delta-/ratio-tracking estimators.

We generate a synthetic dataset composed of volumetric cloud scenes rendered under diverse lighting and medium configurations. Environment illumination is generated using 1000 HDR-like maps synthesized with Stable Diffusion \cite{rombach2022high}, covering a wide range of natural and stylized lighting conditions. In addition, we consider $500$ volumes of cloud formations from the Cloudy Project dataset \cite{leonard2025light}. In total we synthesize $22000$ instances with $4096$ samples per pixel (spp) for the images.

Scenes are rendered with randomized camera poses, light directions, phase function anisotropy parameters, and scattering albedos. The latter are sampled in the high-scattering regime and may vary per color channel, enabling colored volumetric effects. For each instance we generate the scattered radiance from the environment, the scattered radiance from the light direction, the transmittance, and background. We render the images for a normalized (mean = $1.0$) environment and a unitary light intensity since we can combine them with different scalings to cover a wider range of options.

\subsection{Training}

All models are trained using AdamW with identical hyperparameters for fair comparison between the CNN \cite{hofmann2021volumerecon} and HaloRT. We use a learning rate exponentially decaying over $240$K optimization steps from $10^{-3}$ to $10^{-6}$, then held constant for $120$K extra steps to let the network fully converge at a low, stable rate.

The optimization objective combines an $\ell_1$ reconstruction loss with a $\ell_1$
of a logarithmic compression variant to better balance low- and high-intensity regions:
\begin{align*}
    c(x) &= \mathrm{sgn}(x)\,\log(1+(100-1)|x|)/\log 100\\
    \mathcal{L} &= 0.6 \| \hat{I} - I \|_1 + 0.4 \| c(\hat{I}) - c(I) \|_1
\end{align*}

Each step draws a batch of $16$ instances from an in-memory, periodically refreshed training pool, split into micro-batches of $4$ for the forward pass. For every micro-batch we accumulate gradients over a horizontal flip augmentation and, independently, over all three color channels of the spectral-independent network, backpropagating each of the resulting six passes separately with the loss rescaled accordingly. Each training instance stores several independently sampled noisy realizations of its stochastic conditioning feature (next-event-estimation sample for the direct branch, environment sample for the environment branch); one realization is drawn at random per instance at every step, so the same deterministic target is seen under different noise draws across training. The number of environment samples varies uniformly between $1$ and $16$ to make the model agnostic to different noise levels.

\begin{figure}
    \centering
    \includegraphics[width=\linewidth]{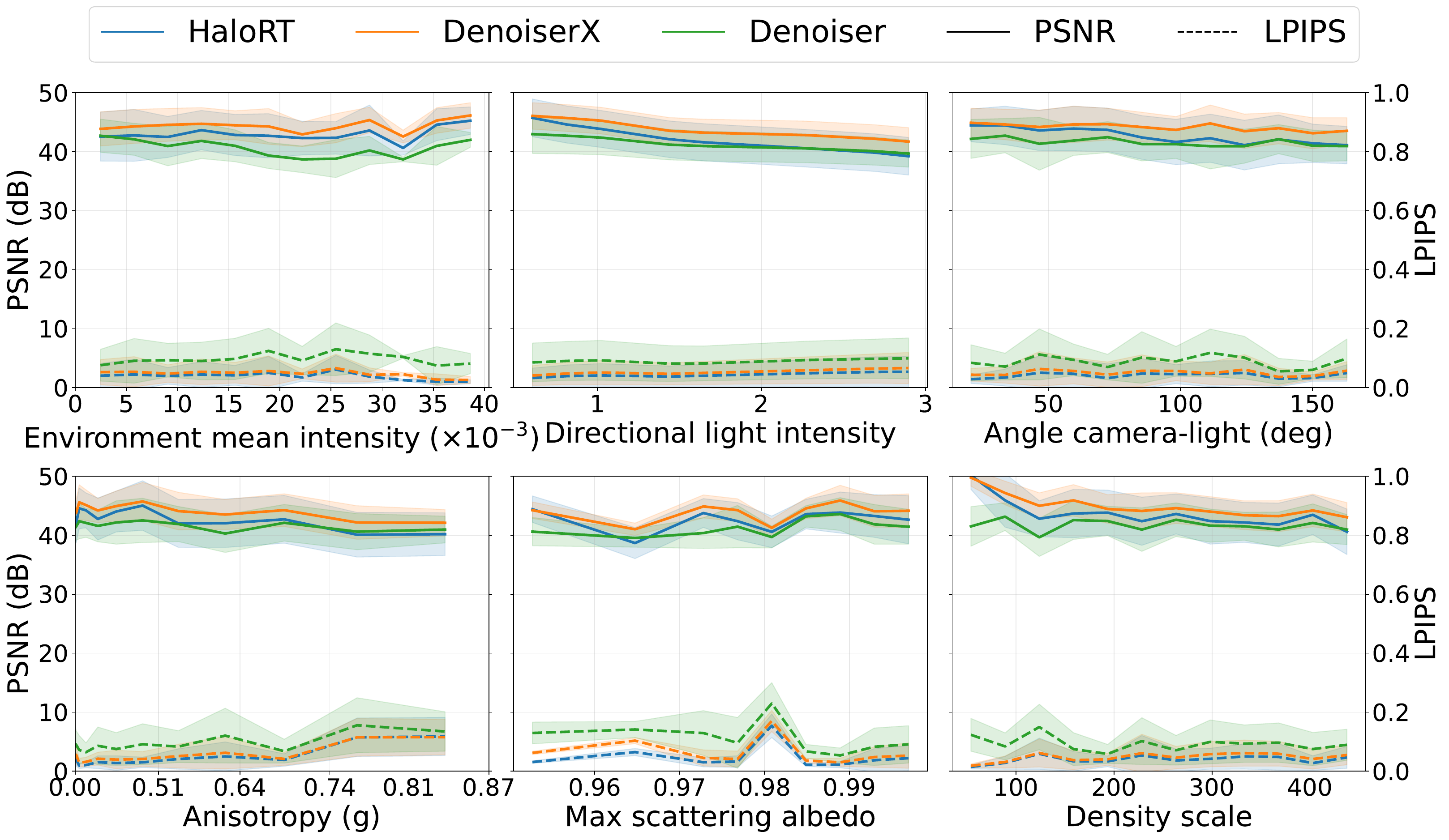}
    \caption{PSNR and LPIPS across the evaluation dataset sorted according to different scene parameters. Our method achieves superior LPIPS although PSNR penalizes global biases not present in denoisers.}
    \label{fig:ssim_comparison}
\end{figure}

\section{Results}

We evaluate three reconstruction approaches under identical rendering and evaluation conditions. The first baseline (\textbf{Denoiser}) is the CNN-based denoiser with $4$-spp Monte Carlo radiance buffer proposed by Hofmann et al. \cite{Hofmann2020neural, hofmann2021volumerecon}, which operates on noisy radiance estimates and statistical features. The second variant (\textbf{DenoiserX}) uses the same architecture but replaces the noisy features with our deterministic features, while retaining the $4$-spp radiance buffer. Finally, our attention-based U-Net (\textbf{HaloRT}) operates exclusively on auxiliary transport features, without requiring path-traced radiance samples.

\emph{Experiment Setup.} All experiments are conducted on an NVIDIA GeForce RTX 5090 GPU. We evaluate all methods on a held-out set of $200$ test images ($512\times512$) rendered for 15 left-out clouds never used for training. We expand the evaluation dataset by including additional environment illumination scales ($\times 0.2$) and directional light intensities ($\times 3$), as well as five distinct directional light colors.

\emph {Quantitative Results.}
We use Structural Similarity Index Measure (SSIM) \cite{wang2004}, Peak Signal-to-Noise Ratio (PSNR), Mean Absolute Error (MAE), Root Mean Square of the pixel-wise predictive Standard Deviation (RMS-SD) as a measure of stability, and Learned Perceptual Image Patch Similarity (LPIPS) \cite{zhang2018unreasonable} for quantitative results (Table~\ref{tab:metrics}). Our method surpasses denoisers in all metrics except PSNR, more notably in stability and time.

We further organize the evaluation dataset according to lighting conditions and volume settings to examine how these parameters affect the performance of the three methods. Fig.~\ref{fig:ssim_comparison} presents the corresponding metrics across the different configurations, allowing performance trends and sensitivities to be compared. While DenoiserX consistently achieves higher PSNR values than our method, both methods exhibit similar trends as the scene parameters vary. In contrast, HaloRT consistently achieves lower LPIPS scores, indicating better perceptual similarity to the reference images across all evaluated configurations. This suggests that, despite the lower pixel-wise fidelity measured by PSNR, our method better preserves perceptually relevant details and light transport characteristics across a wide range of scene conditions.

\begin{figure}
    \centering
    \includegraphics[width=\linewidth]{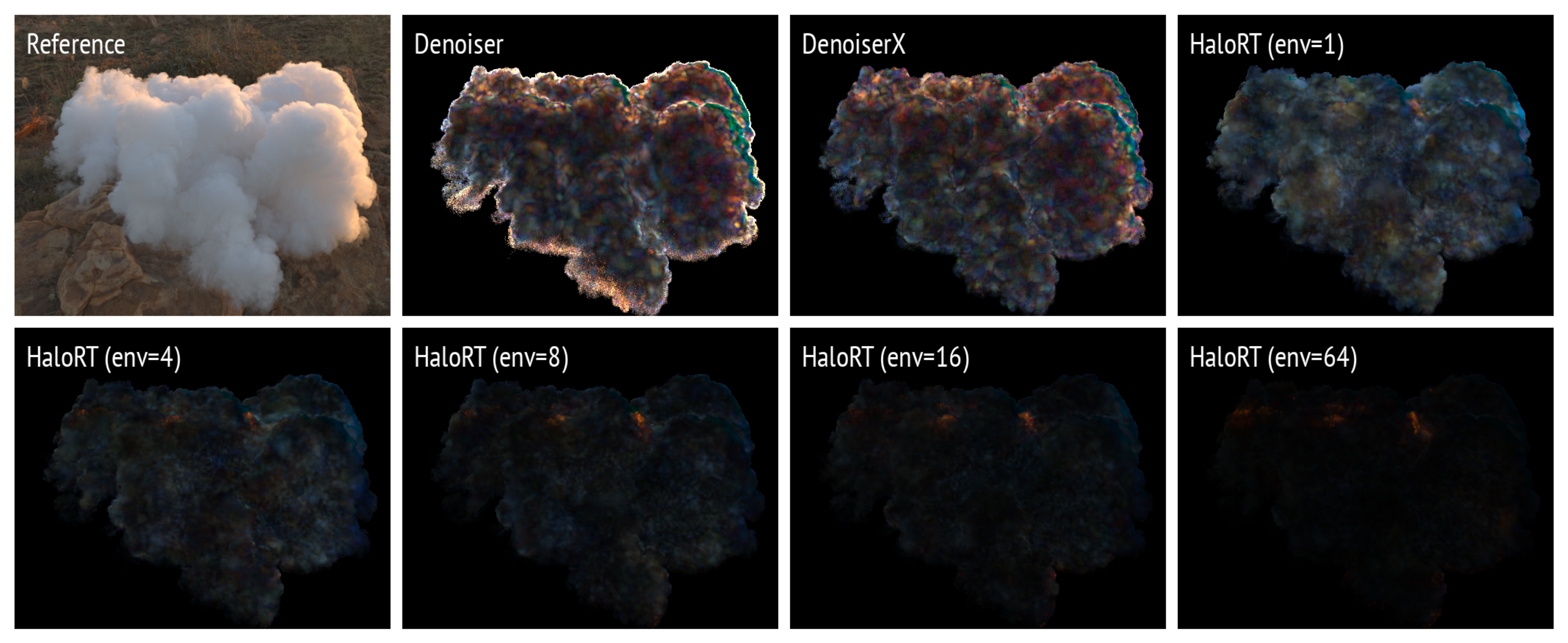}
    \caption{Comparison of stability across $10$ realizations, using maximum absolute pixel-wise deviation from the mean image, amplified by a factor of $16$. Denoisers exhibit higher variability due to stochastic radiance inputs. HaloRT exhibits near-zero deviation with only $8$ environment samples.}
    \label{fig:stability}
\end{figure}

\begin{figure}
    \centering
    \includegraphics[height=.7\linewidth]{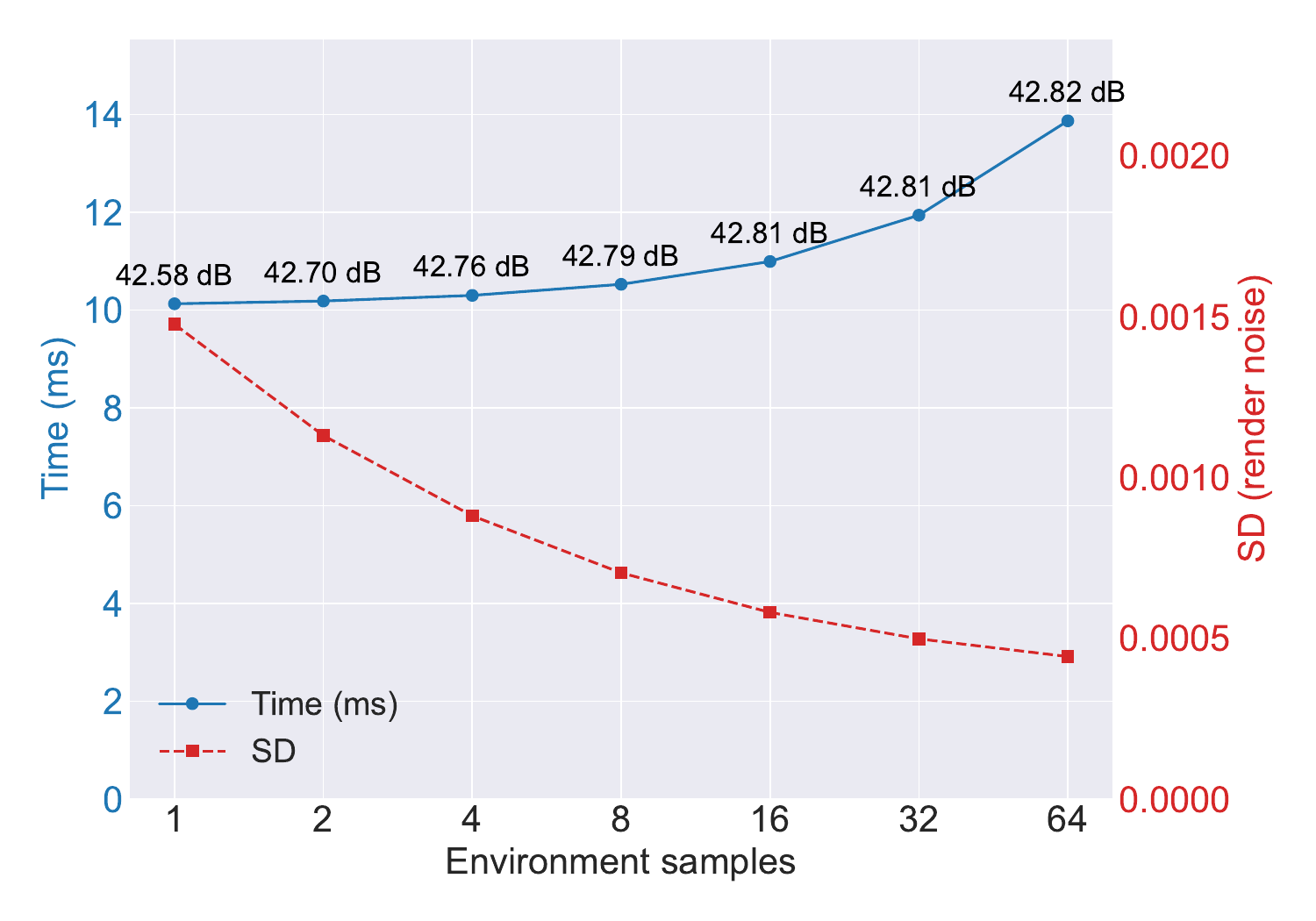}
    \caption{Effect of the number of environment samples on generation time, reconstruction accuracy, and stability, averaged over the evaluation dataset with varying environment lighting.}
    \label{fig:env_samples}
\end{figure}

\begin{figure}
    \centering
    \includegraphics[width=\linewidth]{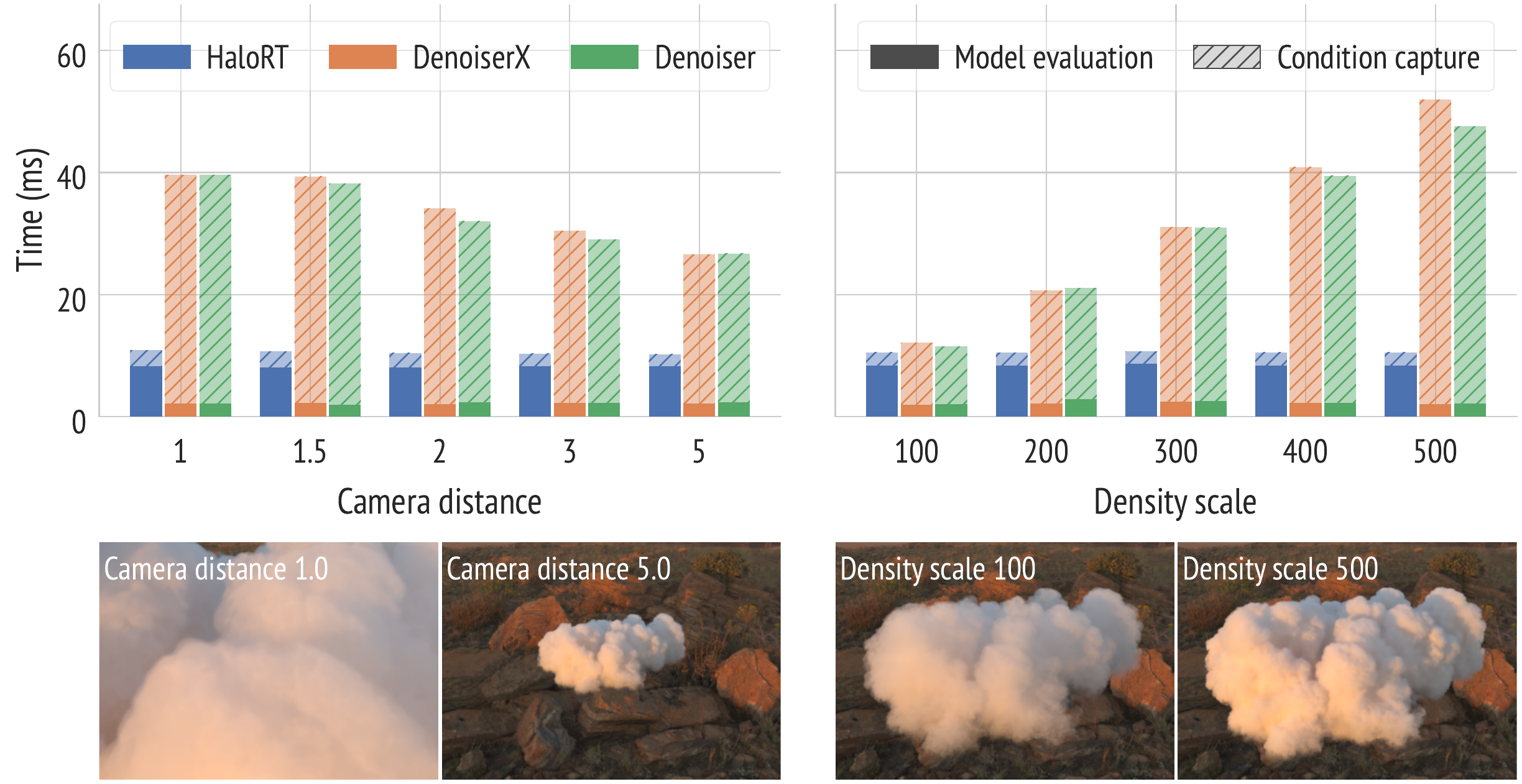}
    \caption{Reconstruction times at different density scales and camera distances. Our network requires more inference time, but since our method does not depend on path sampling, it shows overall better scalability.
    }
    \label{fig:times}
\end{figure}

\emph{Stability.} One of the key properties of our method is that the decoder is largely invariant to feature noise, meaning that stochasticity in the features does not propagate through the model. To demonstrate this, we measure the model's stability as the standard deviation of the inferred image over a given number of realizations (Fig.~\ref{fig:stability}). In particular, we show that only a few environment samples are needed to make the variation imperceptible, with additional samples providing negligible benefits (Fig.~\ref{fig:env_samples}).

\emph{Performance analysis.} Although the CNN used for classical denoising is computationally inexpensive, the radiance estimate it requires -- computed using a few path traces through the volume -- scales poorly with increasing cloud density (Fig.~\ref{fig:times}), as scattering events become more likely and require spawning additional rays. In contrast, since no paths are computed our runtime remains stable.
Along primary rays and environment samples the workload is kept low, due to optimizations like empty-space skipping and fast ratio tracking in sparse volumes, and early termination once the optical depth saturates, along with Russian roulette in ratio tracking, in dense volumes.

As the camera approaches the cloud, denoiser runtimes increase with screen coverage and, thus, with the number of paths to be evaluated. Our inexpensive auxiliary features in combination with the constant computational complexity of the inference step maintain stable runtimes and nearly constant FPS  (Fig.~\ref{fig:times}).

\begin{figure*}
    \centering
    \includegraphics[width=0.95\linewidth]{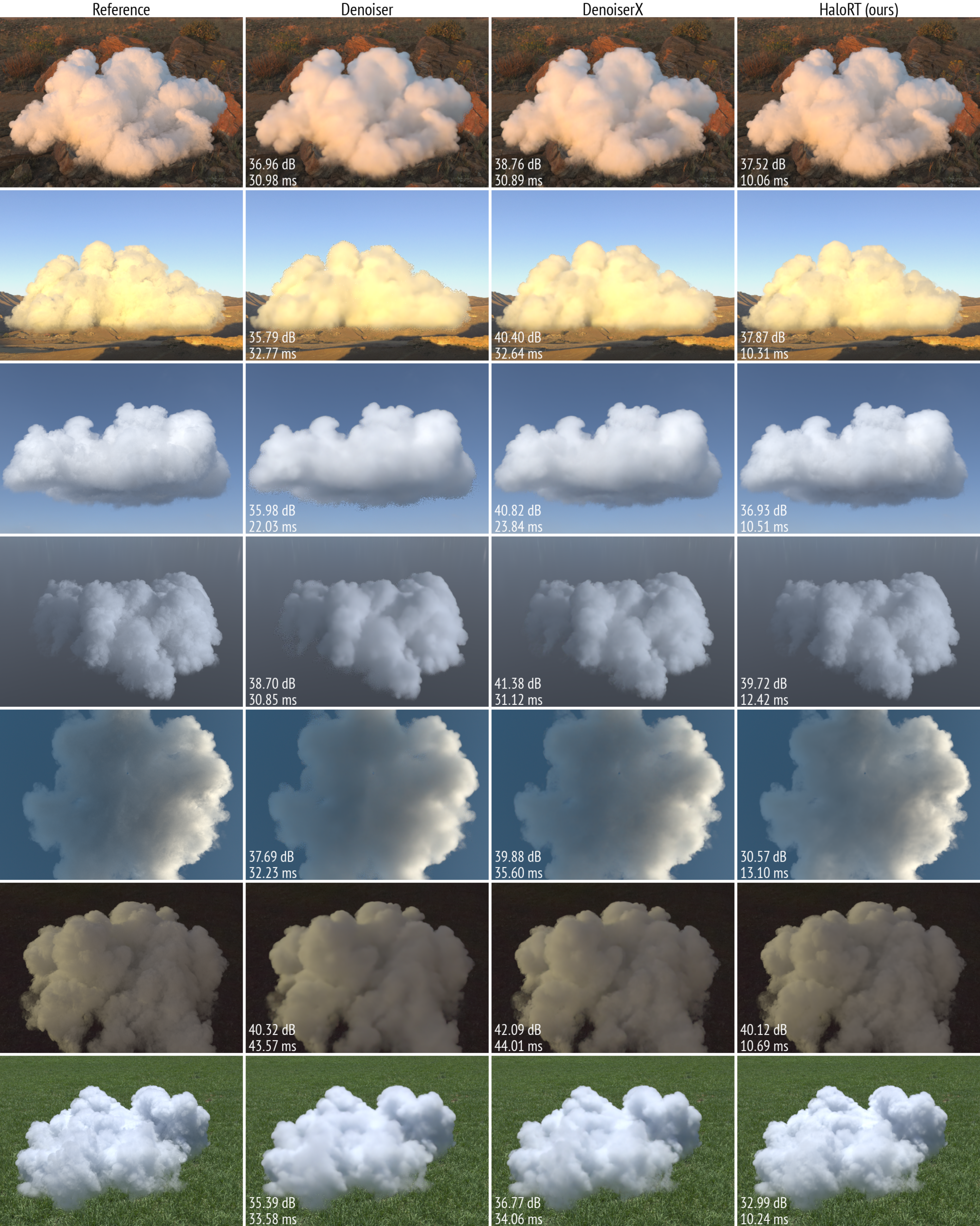}
    \caption{Qualitative comparison of HaloRT against baseline denoisers. Although PSNR is strongly affected by intensity biases, HaloRT better preserves fine details and captures important light transport effects.}
    \label{fig:qualitative_results}
\end{figure*}

\emph{Qualitative Results.} Fig.~\ref{fig:qualitative_results} compares all three different variants qualitatively.

\begin{figure*}
    \centering
    \includegraphics[width=0.95\linewidth]{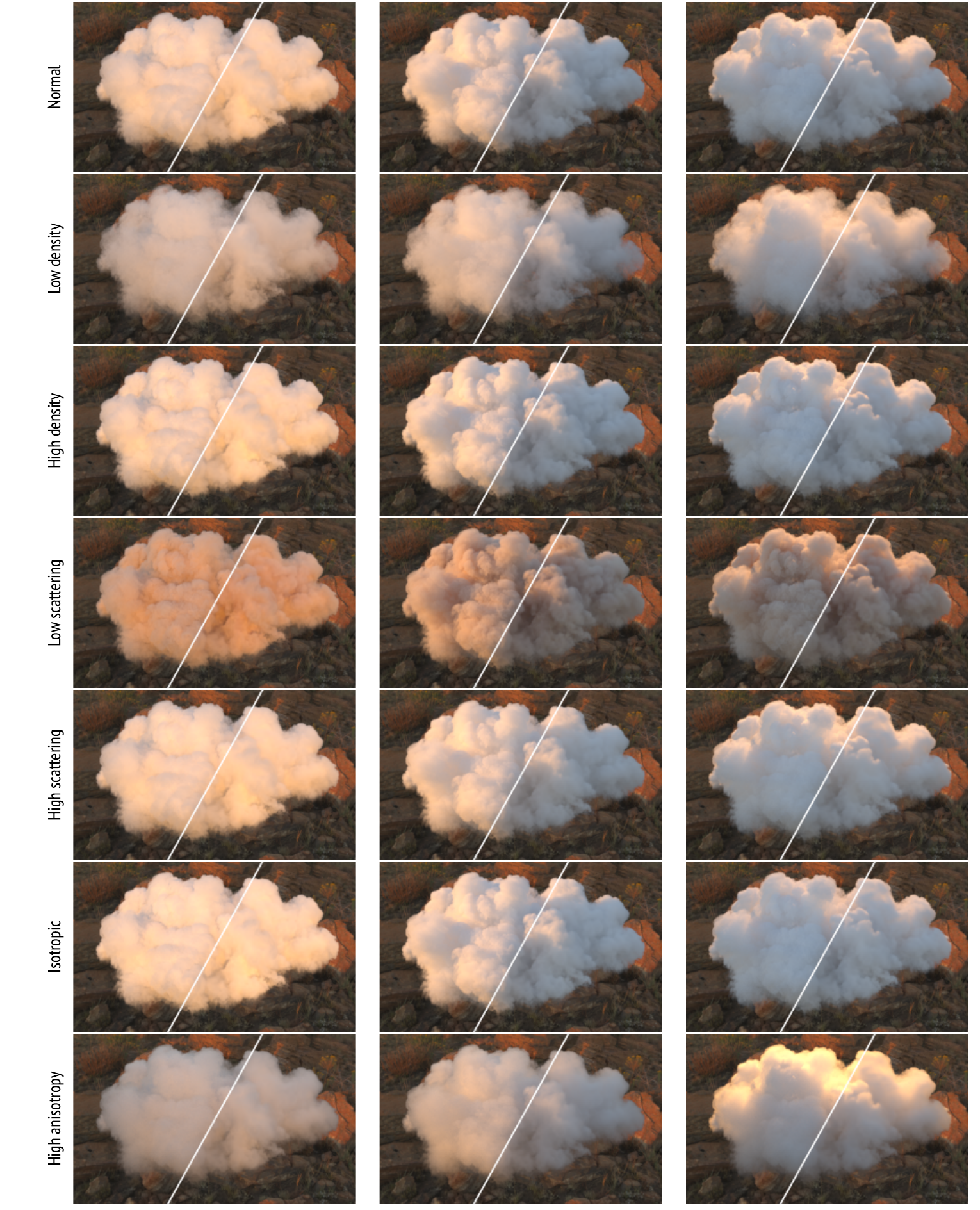}
    \caption{Qualitative comparison. Each pair shows the path-traced reference (left) and HaloRT (right), across variations in density, scattering albedo, and anisotropy under different lighting directions.}
    \label{fig:corner_cases}
\end{figure*}

Fig.~\ref{fig:corner_cases} compares our reconstructions with the path-traced references for a cloud lit from the front, the side, and the back. Starting from a base configuration, each row varies a single medium parameter: density, single-scattering albedo, and phase-function anisotropy. Our method reproduces the brightness and color of the multiply scattered light in all configurations, including the darker appearance of low-albedo media and the bright rim of strongly forward-scattering media under back lighting.

\begin{figure}[ht]
    \centering
    \includegraphics[width=\linewidth]{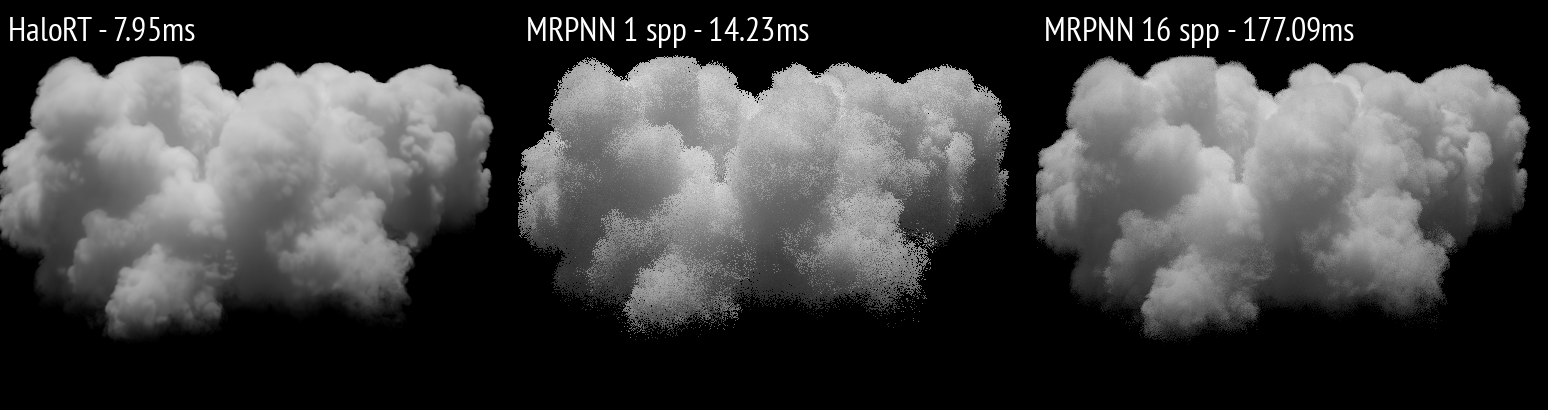}
    \caption{
    Using only one directional monochromatic light, MRPNN \cite{Hu2023MRPNN} 
    shows clearly visible noise at the boundaries. HaloRT's inference
    is noise-free and executes faster than a single pass of MRPNN.
    }
    \label{fig:halo_vs_mrpnn}
\end{figure}

\emph{Directional Light Inference.} In a final experiment, we compare our approach to MRPNN \cite{Hu2023MRPNN} using a single directional light. The experiment is carried out on an NVIDIA GeForce RTX 3090, since the MRPPN code does not execute on the RTX 5090. MRPNN is not designed to efficiently infer the radiance transport in an ambient light environment (see Fig.~\ref{fig:sota_comparison}), but it also requires multiple, yet significantly less samples to generate a noise-free image in the presence of a single directional light (Fig.~\ref{fig:halo_vs_mrpnn}). Ours infers directional light transport with a single pass, producing an accurate, noise-free estimation that runs faster than a single pass of MRPNN.

\section{Discussion and Limitations}

\emph{Feature sufficiency.}
Our results support the view that, for clouds, inexpensive primary-ray features with additional environment samples carry most of the information needed to infer the full transport solution. HaloRT, which never observes a path-traced radiance sample, achieves better perceptual quality and stability than volumetric denoisers at a slightly lower PSNR (Table~\ref{tab:metrics}). The same features also improve the volumetric denoiser substantially when they replace its original auxiliary buffers (DenoiserX vs.\ Denoiser), which indicates that they encode transport information that a radiance-based reconstruction can also exploit.

Even in scenarios which are challenging for HaloRT, i.e., under normal environment lighting where global transport dominates and strong directional light, sufficiency (A1) of our proposed auxiliary feature set still holds reasonably well in practice. In these scenarios, the radiance samples used by a denoiser provide information that our features do not explicitly capture, i.e., the light transport depends on the volumetric structures along paths that leave the vicinity of the primary ray. Since (A1) is only required for plausible scenes $\theta \sim p(\theta)$, the decoder resolves the remaining ambiguities with regularities learned from the training clouds, and its accuracy is therefore tied to how well the training distribution covers the scenes encountered at inference time.

As shown in Fig.~\ref{fig:qualitative_results}, as well as in Fig.~\ref{fig:ssim_comparison}, in the aforementioned scenarios HaloRT shows overall lower PSNR than DenoiserX. However, it is also seen that HaloRT can nevertheless recover features at far more detail, resulting in locally improved and overall tighter reconstructions.

\emph{Noise invariance.}
Since most features are deterministic and only the light-visibility features are estimated stochastically, the decoder becomes nearly invariant to the feature noise (A2). The standard deviation across realizations is less than a quarter of that of DenoiserX and one ninth of that of the Denoiser (Table~\ref{tab:metrics}). A comparison of temporal consistency between the methods can be seen in the supplementary video.

\emph{Cost.}
Unlike path-based reconstruction, our runtime remains stable across densities (Fig.~\ref{fig:times}). In addition, the network dominates the cost: with local attention it scales close to a convolutional network, but inference still takes about $10.5$\,ms per pass at $2048^2$ (Table~\ref{tab:arch_scaling}), and one pass is required per illumination component and color channel.

\emph{Scope.}
We trained and evaluated our method only on cloud-like media from a single dataset, with the Henyey--Greenstein phase function, non-emissive media, and a single volume under distant illumination, consisting of a smooth environment map and one dominant directional light. Surfaces inside or around the medium, local light sources, and emission are not handled. We share the extinction across color channels. Channel-dependent extinction is supported by the formulation but requires one volume traversal per channel. Finally, our training environments are synthesized, and a larger set of captured HDR environments would provide a stronger test of generalization.

\section{Conclusion and Future Work}

We present a real-time method for volumetric light transport  that removes path-traced radiance samples from inference entirely. Rather than denoising stochastic estimates of the image, we reconstruct full multiple scattering from a compact set of auxiliary features: optical depth, multi-scale interaction depths, transmittance towards the light, environment illumination at the expected first interaction, and local medium properties. We formalize this as deterministic decoding under two testable assumptions, feature sufficiency and noise-invariant decoding. We show that MC denoising is a special case in which sufficiency holds trivially and all of the difficulty lies in noise invariance. Our features shift this balance: noise invariance becomes easy to satisfy, while sufficiency holds approximately for clouds. By exploiting the linearity of light transport in the illumination and the independence of color channels, we learn the decoder as two single-channel HaloRT models, which use local self-attention to exchange information across the image at a cost that scales to high resolutions. On unseen clouds, our method outperforms volumetric denoisers in accuracy and stability, runs more than twice as fast than the feature-enhanced denoiser, and depends much less on medium density and camera distance.

Several directions remain open. Extending the features and training data to other media, such as smoke, fire, or subsurface scattering, and to scenes that combine media with surfaces and local lights would broaden applicability and test the limits of the sufficiency assumption. Temporal recurrence or cross-frame attention could improve coherence under camera motion without reintroducing a dependence on path sampling.

The remaining error has three sources, each leaving room for improvements. First, the features are only approximately sufficient (A1). Optimizing the hand-designed feature operators jointly with the decoder could narrow this gap. Second, the learned decoder $S_\Psi$ is limited by model capacity, training data, and optimization, which larger and more diverse datasets and architectures could reduce. Third, the visibility features remain stochastic (A2), and lower-variance estimators, such as deep shadow maps for the directional light, could reduce this residual noise at little cost.

\bibliographystyle{eg-alpha-doi} 
\bibliography{egbibsample}  

\end{document}